\documentclass[fleqn,twoside]{article}
\usepackage{espcrc2}

\usepackage{graphicx}
\usepackage{amssymb}

\newcommand{\AmS}{{\protect\the\textfont2
  A\kern-.1667em\lower.5ex\hbox{M}\kern-.125emS}}

\def\Tr{{\rm Tr}\,}

\title{An Approach to Higher Dimensional Theories Based on 
Lattice Gauge Theory}

\author{M. Murata\address[NIIGATA]{Department of Physics, 
        Niigata University, 
        Ikarashi 2-8050, Niigata 950-2181, Japan}%
              and
        H. So\addressmark[NIIGATA]\thanks{Talk presented by H. So. This work was
        supported in part by Grants-in-Aid for Scientific Research No. 
        and 13135209 from the Japan Society for the Promotion of Science.}}
       
\begin{document}

\begin{abstract}
A higher dimensional lattice space can be  
decomposed into a number of four-dimensional 
lattices called as layers.
The  higher dimensional gauge theory on the lattice can be interpreted as
four-dimensional gauge theories  on the multi-layer
with interactions between  neighboring layers.
We propose the new possibility to realize   the continuum limit of
a five-dimensional  theory based on the property of the phase diagram.
\vspace{1pc}
\end{abstract}

\maketitle

\section{MOTIVATION}
In constructing a  higher dimensional quantum field theory, 
the regularization and the continuum limit are two important keys. 
Particularly, the problem of the limit associates with hard difficulty  
in the higher dimensional case. 
Statistical mechanics usually insists  that  critical behaviors of the phase 
transition  in the theory  are equivalent to those of a mean field theory, 
which has  only a trivial  fixed point.

Many pioneering works on lattice gauge theories were 
trying to overcome the difficulty \cite{creutz,lang,kawai,ejiri1,ejiri2}, 
although the continuum limit is not taken strictly. 
Where is the continuum limit?  
By using well-known  4-dimensional theories such as QED and QCD, 
is it possible to construct the continuum limit by 
the   related critical behavior near the critical point?   
Our purpose   is to construct  a $D$-dimensional pure Yang-Mills theory 
by arranging   a  number of  4-dimensional Yang-Mills theories with 
appropriate couplings.

\section{OUR APPROACH}
\begin{figure}[htb]
\includegraphics[width=15pc]{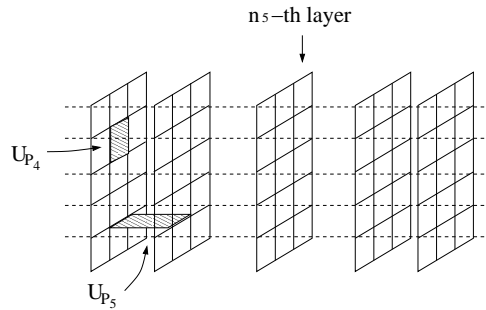}
\caption{Layer decomposition of 5-dimensional lattice space.}
\label{fig:layer}
\end{figure}

A $D$-dimensional lattice space is decomposed into 
$N_D$ layers with 4 dimensions like as  Fig.\ref{fig:layer}.
Originally, Fu and Nielsen  investigated the system 
to show dynamical dimensional reduction 
by using the characteristic vacuum which confines for 
extra dimensional directions and deconfines for 4 dimensions  
called a layer phase \cite{F-N}. 
We do not assume that the layer phase exists, but use only 
the decomposition. 

From now, we focus on a 5-dimensional theory to study the possibility of 
the construction for higher dimensional field theories explicitly but 
an extension to general dimensions shall be mentioned in the final section. 
Our starting action for $SU(2)$ gauge group is written as 

\begin{equation}
S_{\rm lat}
=\frac{\beta_{4}}{2}\sum_{P_4}\left[2-\Tr U_{P_4}\right]
  +\frac{\beta_{5}}{2}\sum_{P_{5}}\left[2-\Tr U_{P_{5}}\right], 
\label{lact}
\end{equation}
\noindent
where $\beta_{4}$ is proportional to a usual 4-dimensional coupling constant
 inside a layer 
and  $\beta_5$ is  a coupling constant between neighboring layers. 
$P_4$ implies a plaquette inside a layer and $P_5$ does a plaquette between 
neighboring layers.

\section{PHASE STRUCTURE}
\begin{figure}[htb]
\includegraphics[width=12pc]{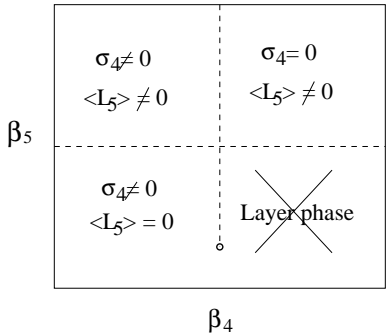}
\caption{Expected phase diagram. Layer phase can not realize in this system. }
\label{fig:phase1}
\end{figure}

To display  the phase diagram explicitly, two order parameters are introduced; 
(1) Creutz ratio ($\sigma_4$) for 4-dimensional Wilson loops  
and (2) 5-dimensional Polyakov loop ($\langle L_5\rangle$).
Since a theory with $\beta_5=0$ is equivalent to a  pure 4-dimensional 
Yang-Mills theory which is one phase, we expect a phase structure such as 
Fig.\ref{fig:phase1}.

\begin{figure}[htb]
\includegraphics[width=17pc]{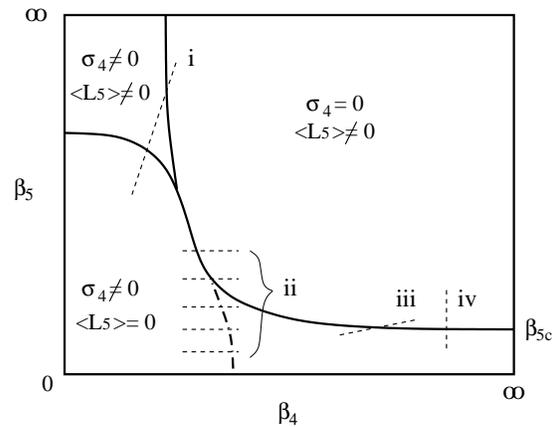}
\hspace{-1.5 cm}
\caption{Numerical phase diagram. The vertical dashed line represents crossover,  
not phase transition one.}
\label{fig:phase2}
\end{figure}

We have calculated numerically these parameters for the system and 
obtained the numerical phase diagram in Fig.\ref{fig:phase2}.
In the figure, along line i we can see both  4-dimensional and  5-dimensional  
deconfinement transitions in the large $\beta_5$. 
By wider calculations along  lines ii, we recognize 
the phase transition is of  1st order near  $\beta_4 \approx \beta_5$.  
The results of lines iii and  iv suggest that  
 both  4-dimensional and  5-dimensional  
deconfinement transitions in the small $\beta_5$ are of 2nd or weakly 1st order,  
because we can not  see any hysteresis loop in the coupling region.
This critical point in the $\beta_4 \rightarrow \infty$ 
is noted as $\beta_{5c} (\approx 0.6)$. 
Our surprising remark is that the diagram Fig.\ref{fig:phase2} is not 
quantitatively changed  for various $ N_5 =2,3,5,6,8$ and its stable property  
may help us to take the limit $N_5 \rightarrow \infty$.  
\section{MULTI-LAYER WORLD AND 5-SPACE}
\subsection{Multi-Layer World}
Our  simple consideration finds our way to  existence of  a 4-dimensional continuum 
system with finite inter-layer coupling($\beta_5$).  
Near the critical point ($\beta_5 \approx \beta_{5c}$), 
the inverse ($H$) of correlation length for Polyakov loop is written as 
\begin{eqnarray}
H = m_h a_4 \propto |\beta_5 - \beta_{5c}|^{\nu} ,
\label{higgs}
\end{eqnarray}
\noindent
where the value of  $\beta_{5c}$ is approximately 0.6 
and $\nu$ implies the critical exponent which is 0.5 in 
a mean field theory. 
We call  
the picture that many gauge fields on different layers interacts 
each other with finite coupling as multi-layer world. 
It is noted that link variables with 5-th direction 
behave as bi-fundamental fields with a finite coupling.  

\subsection{5-dimensional space}
\begin{figure}[htb]
\includegraphics[width=15pc]{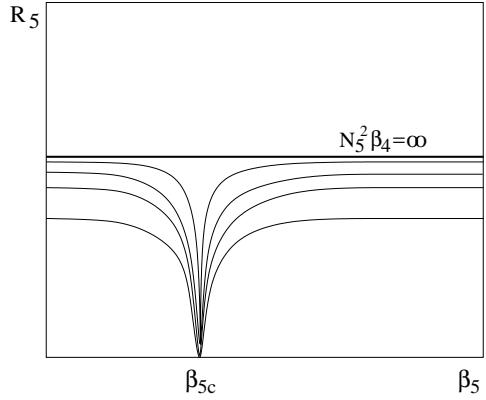}
\hspace{-1.5 cm}
\caption{Curves of continuum limit associated our 
limiting operation.}
\label{fig:env}
\end{figure}

Can we construct a 5-dimensional space not an internal space? 
A straightforward way uses excitation masses $M_{i,j}$ 
between layers  corresponding to Kaluza-Klein(K-K) modes, 
\begin{eqnarray}
M_{ij}^2
&\equiv& \frac{\beta_5 g_4^2 }{4\ a_4^2} \left\{
   \left[
    \begin{array}{ccccc}
    1 & -1 &        & & \\
    -1&  2 & \ddots & &  \\
      & \ddots & \ddots & \ddots & \\
      &    & \ddots & 2 & -1 \\
      &    &        & -1&  1 \\
    \end{array}
    \right] \right. \nonumber  \\ 
   & + & \langle L_5\rangle^2
  \left.\left[
    \begin{array}{ccccc}
    1 &  &     &   &  -1\\
      & 0&     &  0& \\
      &  & \ddots &   &  \\
      & 0&        & 0 &  \\
    -1&  &        &   &  1 \\
    \end{array}
    \right]
   \right\} .
\label{kkmass}
\end{eqnarray}
\noindent
For a lower excited mode with label $k$, the simple formula of the mass 
is obtained as 
\begin{equation}
M_k \sim  
\sqrt{\frac{\beta_5}{N_5 \beta_4}} \left(\frac{2\pi  k}{a_4} \right)\ .
\label{kkm}
\end{equation}
\noindent
In order to remain   finite masses of K-K modes, we must 
keep $\sqrt{N_5 \beta_4} a_4$  finite for large $N_5$ and $\beta_4$ 
with small $a_4$ from Eq.(\ref{kkm}). 
To go through a 5-space from our 4-dimensional multi-layer world, 
we need to balance inter-layer dynamics and inside-layer one, 
\begin{eqnarray}
{\cal R}\equiv \frac{M_k^2}{m_h^2}
&  \sim  & \frac{\beta_5 \lambda_2(N_5)}{\beta_4 H^2(\beta_4,\beta_5,N_5)}   \nonumber \\
 & & \propto \frac{\beta_5}{N_5^2 \beta_4 |\beta_5 - \beta_{5c}|^{2\nu}} \ . 
\label{defcalR}
\end{eqnarray}
From Eq.(\ref{defcalR}), three parameters $(\beta_4, \beta_5, N_5)$ tunings are 
necessary (See Fig.\ref{fig:env}).
Comparing  $M_{\rm K-K} \sim a_5^{-1}$ with Eq.(\ref{kkm}), 
a lattice spacing $a_5$ along 5th dimension can be defined by  
$a_5 \equiv \frac{\sqrt{\beta_4}}{4\sqrt{\beta_5}} a_4$.

\section{SUMMARY AND PROBLEMS}
We summarize main three steps; 
$a)$ To find a second order phase transition in the meaning of 
    4-dimensional statistical mechanics. 
$b)$  To take a 4-dimensional  continuum limit 
     (Multi-Layer World).
$c)$ To compare a 4-dimensional scale with 
     an extra dimensional scale, 
   i.e.  confinement and Kaluza-Klein modes.
For detailed analysis, see Ref.~\cite{M-S}. 

Finally, the following problems remain;
$A)$  Further  detailed study
    for large   $N_5$ and $\beta_4$.
$B)$  To estimate contribution of bi-fundamental field
     for $\sigma_4$.
$C)$  To recover the rotational symmetry relating to 
anisotropy between 4-Scale and 5-Scale.   
$D)$  6 or higher-dimensional extension of these decomposition 
is straightforward but their phase diagram analysis is not so easy.

\end{document}